\documentclass[prl,twocolumn]{revtex4}
\usepackage{epsfig}
\usepackage{bm}

\begin{document}

\title{Hamiltonian distributed chaos in the north-south dipole and quadrupole teleconnections}

\author{A. Bershadskii}

\affiliation{
ICAR, P.O. Box 31155, Jerusalem 91000, Israel
}

\begin{abstract}
  
  It is shown that the north-south teleconnections in the northern hemisphere: North Atlantic (NAO), East Pacific (EPO), Western Pacific (WPO) dipole oscillations and  Pacific/North American quadrupole pattern (PNA), are dominated by the Hamiltonian distributed chaos on the daily to intraseasonal time scales. Differences of the chaotic properties of the dipole and quadrupole oscillations as well as their relation to the surface air temperature have been briefly discussed. A chaotic spectral affinity of the PNA quadrupole pattern to the Arctic Oscillation  and the Greenland blocking phenomenon have been considered in this context.  
  
\end{abstract}

\maketitle

\section{Introduction}

    The four major north-south teleconnections: North Atlantic (NAO), East Pacific (EPO), Western Pacific (WPO) oscillations and  Pacific/North American pattern (PNA) reflect planetary-scale recurring patterns (atmospheric oscillations) of circulation and pressure anomalies over the Atlantic and Pacific oceans in the Northern hemisphere (see, for instance, Refs.\cite{bl}-\cite{cfl} and references therein). While the NAO, EPO, and WPO are north-south dipoles of anomalies spanning mainly over corresponding oceans the Pacific/North American pattern (PNA) is a quadrupole including also the intermountain region of North America (i.e. land together with ocean) and, therefore, is a special case. 
    
    The decadal, annual and seasonal large-scale atmospheric oscillations over Atlantic and Pacific oceans attracted major attention of researchers whereas the statistical properties of their fluctuations on daily to intraseasonal time scales are much less studied. Meanwhile, just on these time scales the surface temperature dynamics, very important from meteorological point of view, is dominated by Hamiltonian distributed chaos \cite{b1}. \\

  The exponential frequency spectrum 
$$
 E(f) \propto \exp-(f/f_c) \eqno{(1)}
$$ 
is often observed in the systems with deterministic chaos \cite{fm}-\cite{b2}. A more complex stretched exponential spectrum is observed for the Hamiltonian chaotic systems
$$
E(f ) \propto \int_0^{\infty} P(f_c) \exp -(f/f_c)~ df_c  \propto \exp-(f/f_0)^{\beta}  \eqno{(2)}
$$
with a distribution $P(f_c)$ of the characteristic frequency $f_c$, and the $\beta = 1/2$ or $3/4$ depending on the boundary conditions \cite{b1}. 

   Naturally, most of the models in the climate theory are Hamiltonian dynamical systems (see Refs. \cite{gl}-\cite{gl2} and references therein). The NAO, EPO, WPO and PNA indices represent the four major patterns of the geopotential height variability over the Atlantic and Pacific oceans. One can expect that power spectra computed for these indices have the characteristic for the Hamiltonian distributed chaos form Eq. (2) with the $\beta =1/2$ or 3/4. 
   
\section{The North Atlantic oscillation} 
   
   In the Northern Hemisphere much of the Atlantic ocean is covered by the 
North Atlantic Oscillation (NAO). The NAO index can be defined as the difference in normalized sea-level pressure anomalies between Southwest Iceland (a northern node) 
and Azores (a southern node). It is so-called station-based method of the NAO index computation (see, for instance, Refs. \cite{hur},\cite{j},\cite{gr} and references therein). There is also another method of the NAO index computation. This method uses gridded climate datasets with empirical orthogonal analysis -  EOF (see, for instance, Refs. \cite{tw1},\cite{tw},\cite{fo} and references therein). There are also different modifications of the above mention methods and, therefore, there are different versions of the NAO index (this is also right for other climate indices). All these methods have their strong and weak sides. \\

     The positive phase of the NAO corresponds to {\it below} normal  pressure and heights across the high latitudes of the North Atlantic and {\it above} normal pressure and heights over the central North Atlantic, the western Europe and eastern United States. The negative phase corresponds to an opposite pattern of pressure and height anomalies. 

   Strong influence of the North Atlantic oscillation was registered  in the surface temperature dynamics from eastern North America to central Europe (and even to the north-central Siberia), and from Greenland and Scandinavia to the Middle East \cite{hu}.  \\ 

  The daily NAO index computed by the station-based method - the difference in normalized sea-level pressure anomalies between Southwest Iceland and Azores \cite{gr}, will be analysed for 1878-2014yy period as an example of the first (station-based) method. All other examples belong to the second method with the NAO, EPO, WPO and PNA daily indices based on the dipole and quadrupole centers of action of 500mb height oscillating patterns \cite{nao}-\cite{pna} (see Ref. \cite{kal} for the details of the NCEP-NCAR R1 reanalysis). To emphasize large-scale properties of the teleconnections the height fields were spectrally truncated at the indices computation (see the Refs. \cite{nao}-\cite{pna} for more technical details). An additional information about the NAO and PNA indices and a different methodology of their computation can be found in the Refs. \cite{nao2} and \cite{pna2}. \\
 
   Figure 1 shows power spectrum of the NAO daily index (station-based) computed for the period 1878-1932yy. The data for the computations were taken from the site \cite{zen}. The maximum entropy method with an optimal resolution \cite{oh} has been used for the computation. The straight line indicates correspondence to the Eq. (2) with $\beta = 1/2$. Figure 2 shows analogous spectrum for the period 1932-2014yy \cite{bz}.\\

\begin{figure} \vspace{-1.5cm}\centering
\epsfig{width=.45\textwidth,file=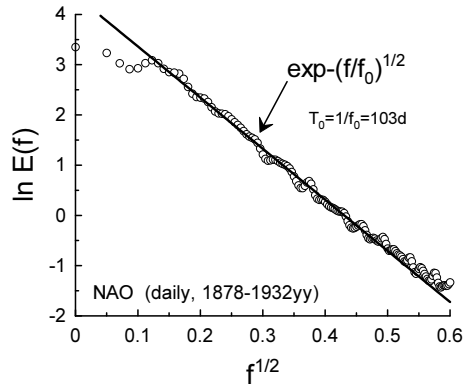} \vspace{-4cm}
\caption{Power spectrum of the NAO station-based index computed for the period 1878-1932yy. The data for the computations were taken from the site \cite{zen}.} 
\end{figure}
   
\begin{figure} \vspace{-1.4cm}\centering
\epsfig{width=.45\textwidth,file=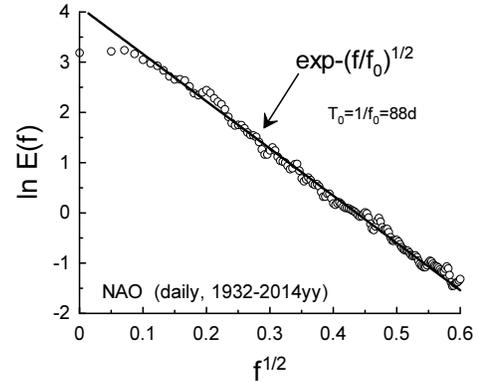} \vspace{-4.1cm}
\caption{As in the Fig. 1 but for the period 1932-2014yy.} 
\end{figure}
   
  Now let us turn to the second method \cite{nao}. The North Atlantic oscillation is a combination of the parts of the West Atlantic and  East Atlantic patterns and represents a north-south dipole of anomalies: one center is located over Greenland and the other center (of opposite sign) covering the central latitudes of the North Atlantic between 35-45N (see Fig. 3).  At the computations of the NAO index the area-weighted mean 500-hPa geopotential height fields of the region 55-70N;70W-10W is subtracted from 35-45N; 70W-10W area averaged region.

\begin{figure} \vspace{+4cm}\centering
\epsfig{width=.85\textwidth,file=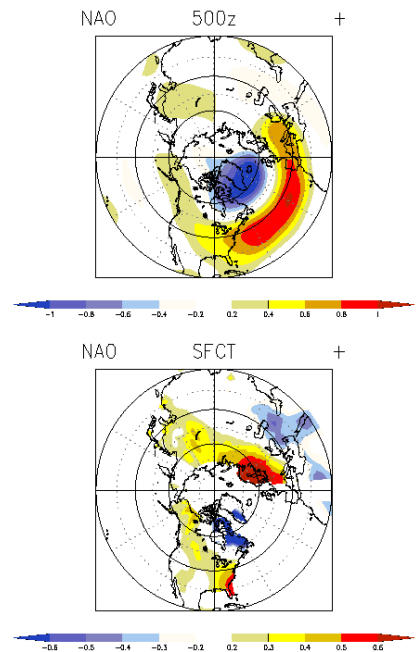} \vspace{-5.5cm}
\caption{Map of the NAO at 500mb height (positive phase), and corresponding surface temperature anomaly over land (the lower picture) \cite{nao}.} 
\end{figure}
\begin{figure} \vspace{-1.2cm}\centering
\epsfig{width=.45\textwidth,file=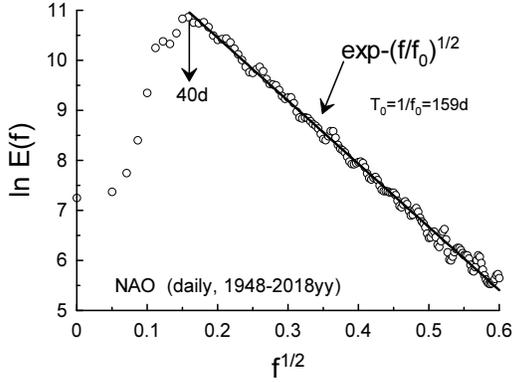} \vspace{-4.3cm}
\caption{Power spectrum for the wavelet regression detrended daily time series corresponding to the NAO index for 1948-2018yy period.} 
\end{figure}
   
\begin{figure} \vspace{+4cm}\centering
\epsfig{width=.85\textwidth,file=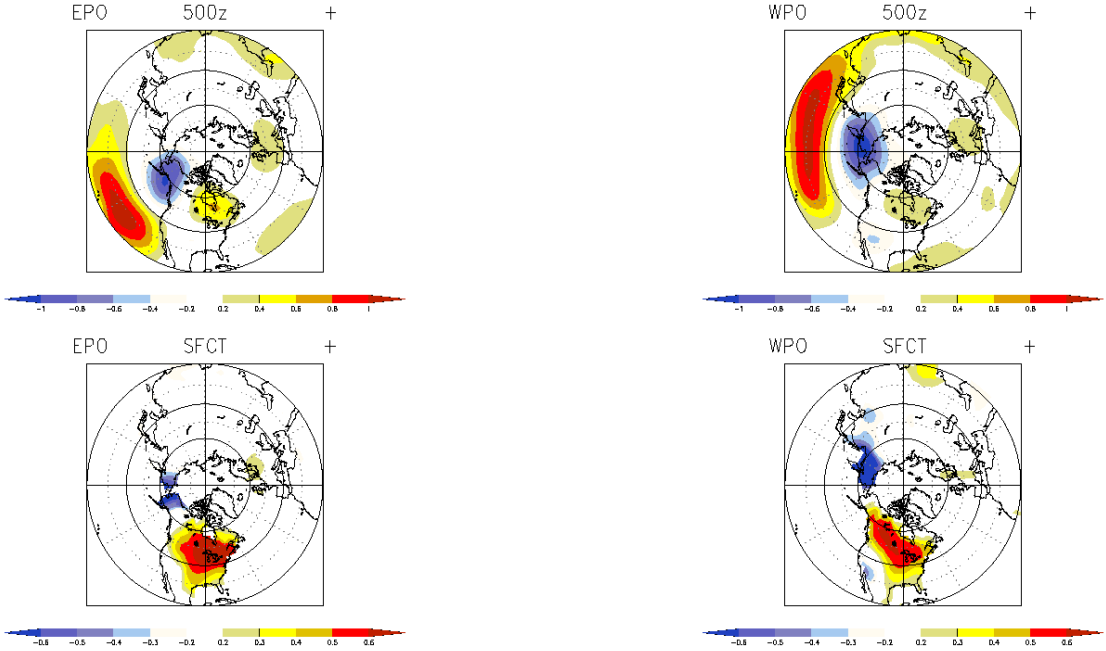} \vspace{-5cm}
\caption{As in Fig. 3 but for the EPO \cite{epo}.} 
\end{figure}
\begin{figure} \vspace{-1.1cm}\centering
\epsfig{width=.45\textwidth,file=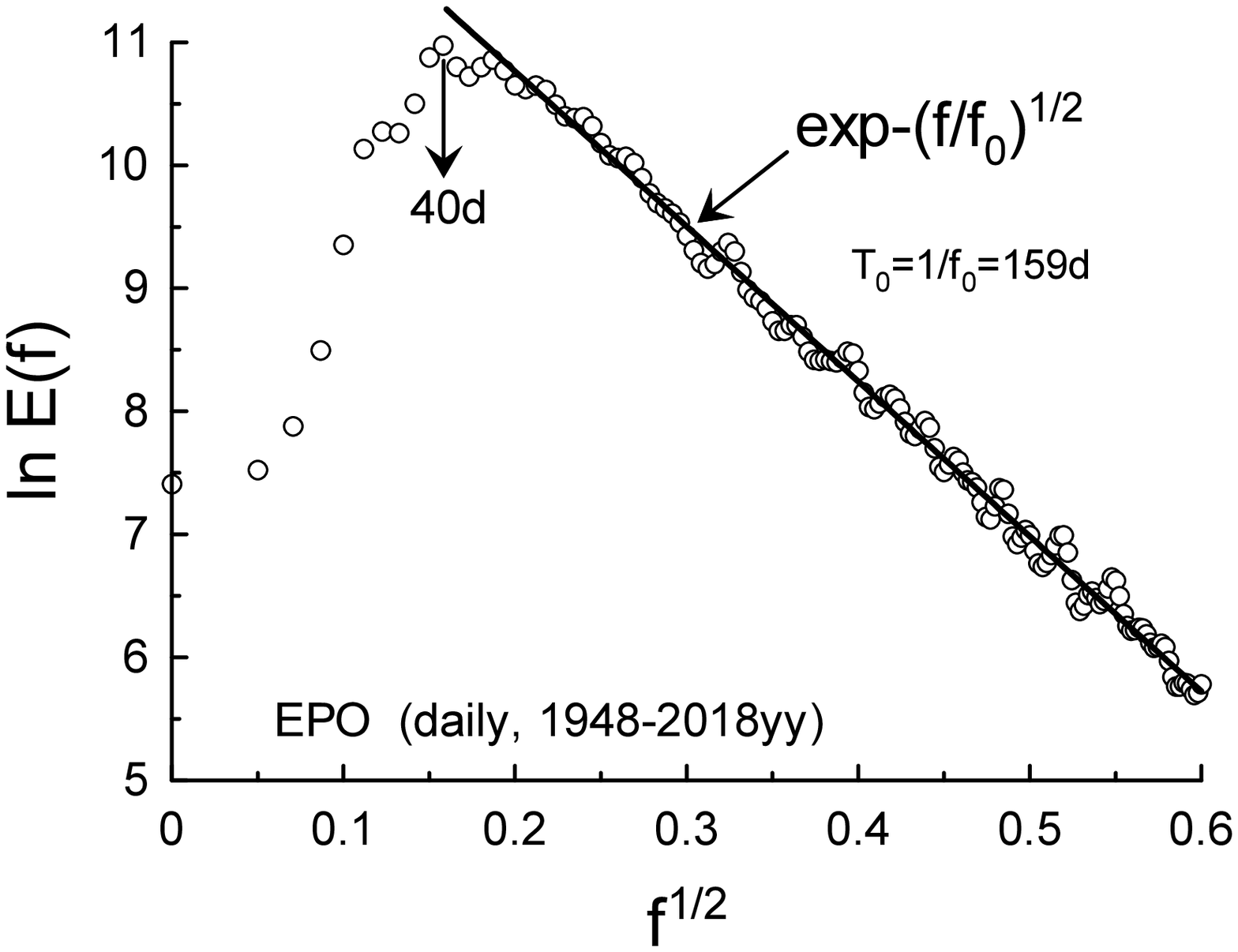} \vspace{-4.4cm}
\caption{As in Fig. 4 but for EPO index.} 
\end{figure}
\begin{figure} \vspace{+4.7cm}\centering
\epsfig{width=.85\textwidth,file=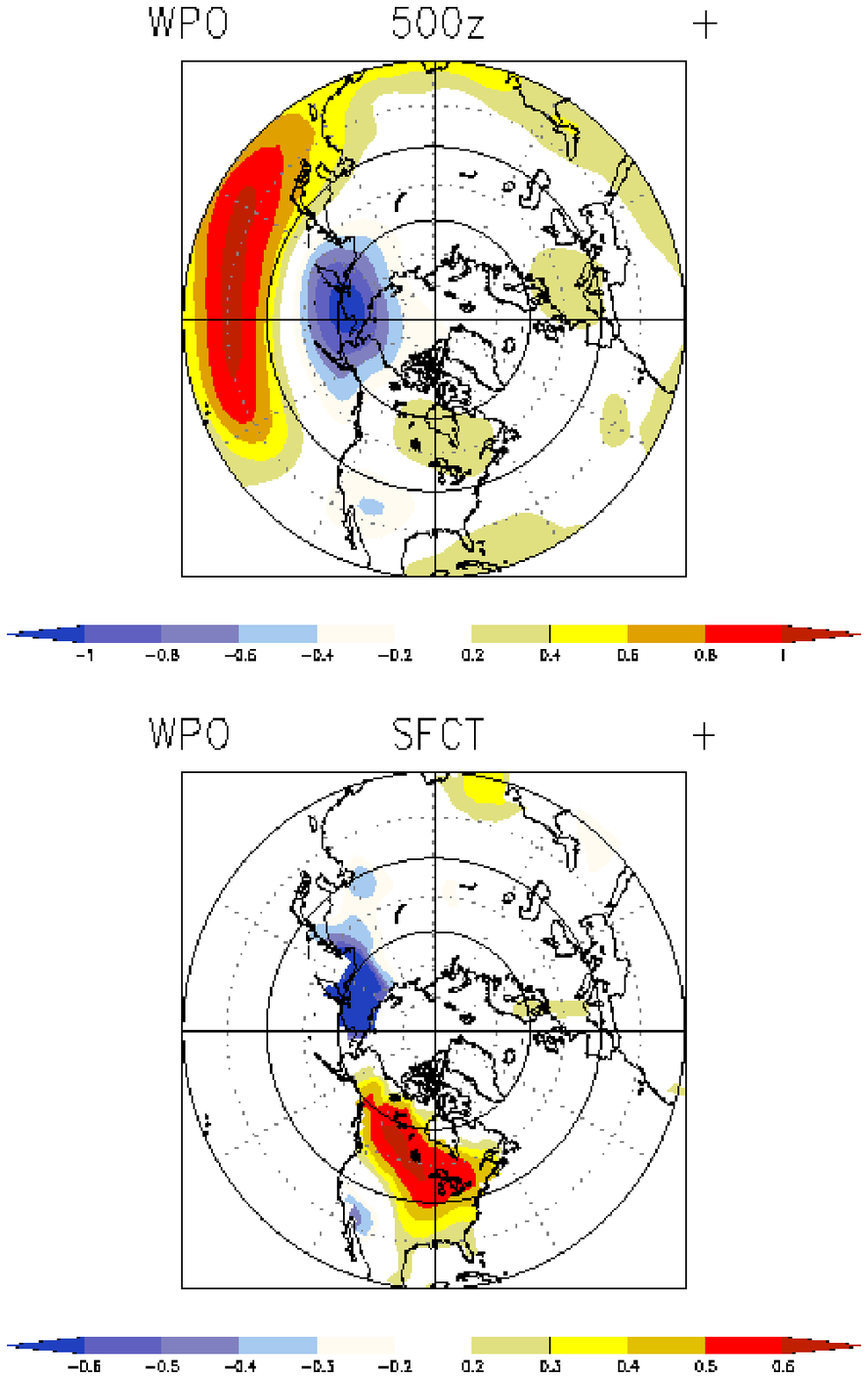} \vspace{-5cm}
\caption{As in Fig. 3 but for the WPO \cite{wpo}.} 
\end{figure}
\begin{figure} \vspace{-1.6cm}\centering
\epsfig{width=.45\textwidth,file=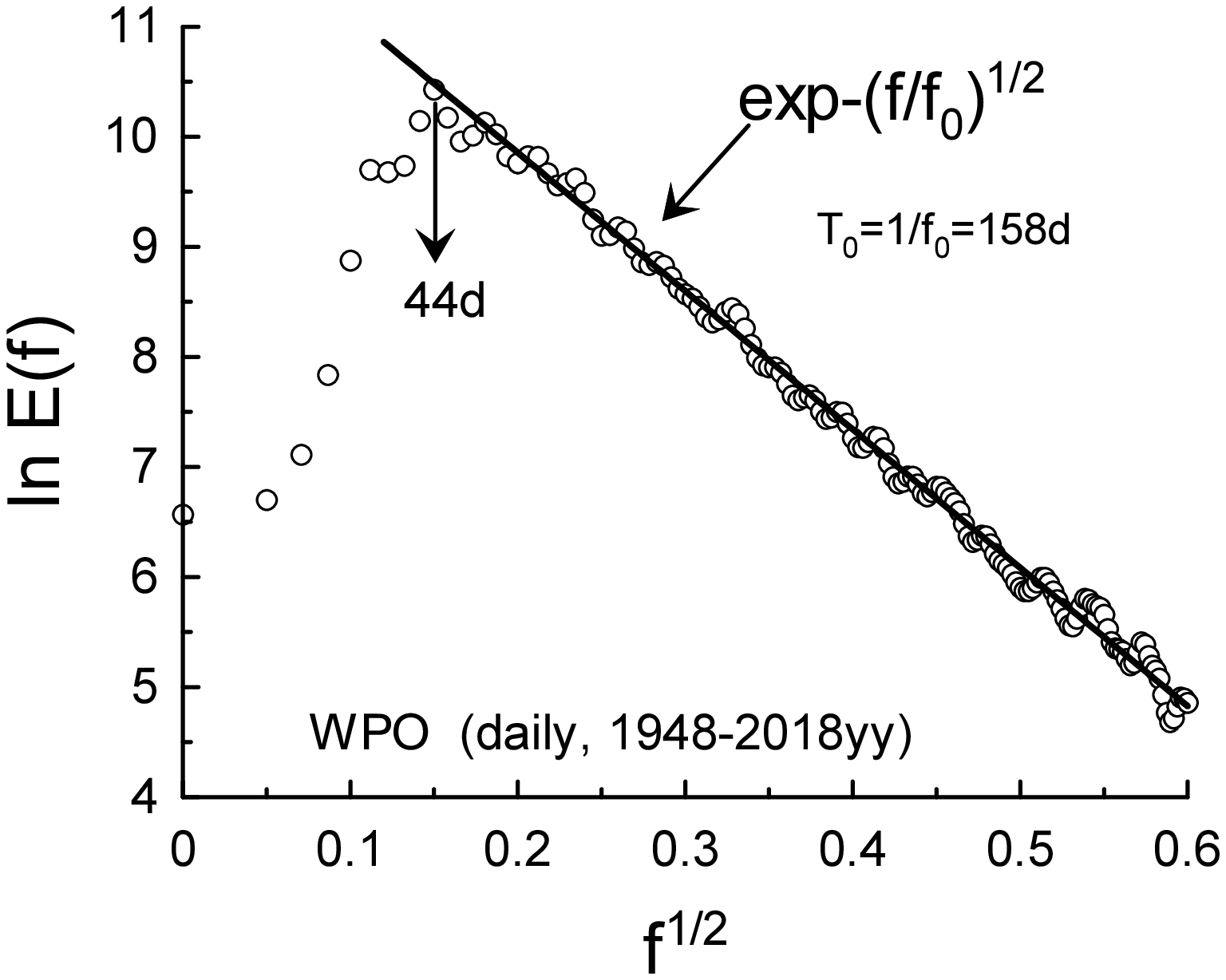} \vspace{-4.04cm}
\caption{As in Fig. 4 but for WPO index.} 
\end{figure}
\begin{figure} \vspace{+4cm}\centering
\epsfig{width=.85\textwidth,file=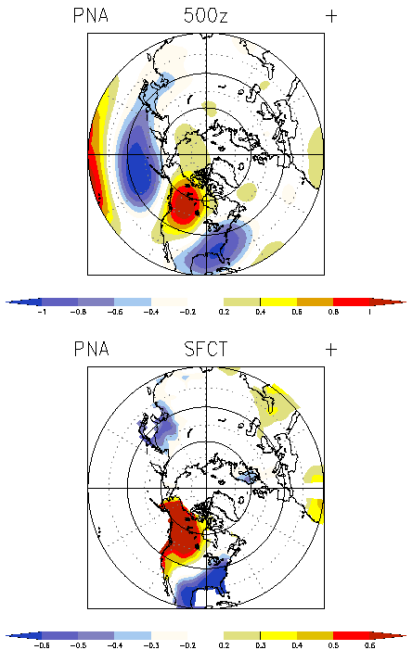} \vspace{-5.5cm}
\caption{As in Fig. 3 but for the PNA \cite{pna}.} 
\end{figure}
\begin{figure} \vspace{-1.55cm}\centering
\epsfig{width=.45\textwidth,file=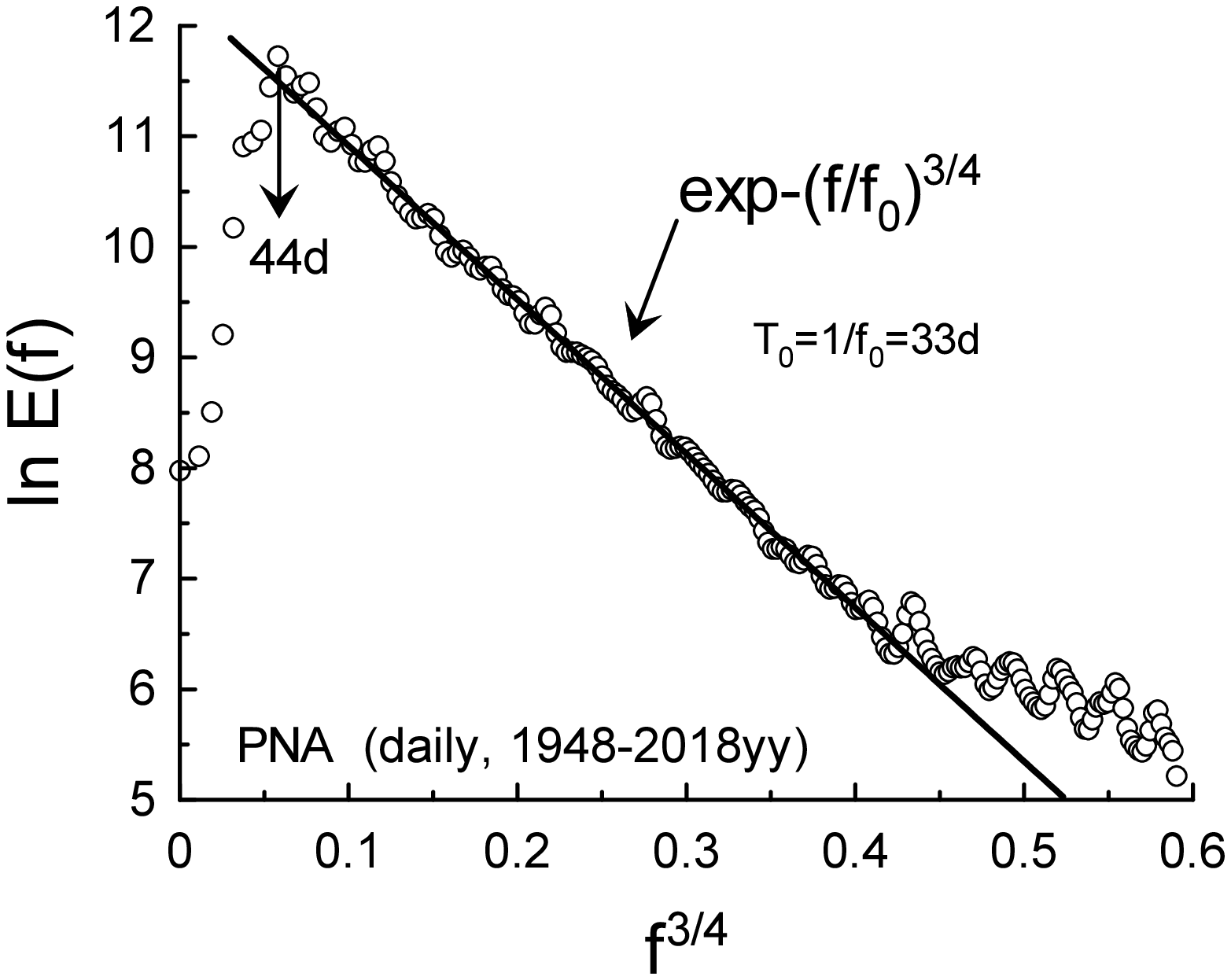} \vspace{-4.05cm}
\caption{As in Fig. 4 but for PNA index.} 
\end{figure}
~\\          
    The daily time series for the NAO index is available at site Ref. \cite{nao} for the 1948-2018yy period. In order to remove the low-frequency annual and seasonal modes a subtracting of a wavelet regression from the daily time series was made using the simplest Haar wavelet \cite{ogd}.  

    Figure 4 shows power spectrum computed for the Haar wavelet detrended daily time series of the NAO index for the 1948-2018yy period (the spectrum was computed by the maximum entropy method \cite{oh}). The straight line in the figure (the best fit) indicates the stretched exponential decay Eq. (2) with the $\beta =1/2$. The fundamental period $T_f \simeq 40$d (indicated by the vertical arrow in the Fig. 4) and $T_0=1/f_0 \simeq 159$d.

\section{East and West Pacific Oscillations}

  A north-south dipole of anomalies, similar to the NAO but located in the eastern Pacific, is known as East Pacific Oscillation (EPO) \cite{bl}. Computations of this index are similar to those used for the NAO index computations but the area averaged region 55-65N;160W-125W was subtracted from the area averaged region 20-35N; 160W-125W in this case \cite{epo}. 

   In the positive phase of the EPO pressures, heights and temperatures are higher to the south and lower to the north while the negative phase corresponds to an opposite pattern (cf. Fig. 5).
   
    Figure 6 shows power spectrum computed for the Haar wavelet detrended daily time series of the EPO index for the 1948-2018yy period (the data were taken from the site Ref. \cite{epo}). The straight line in the figure (the best fit) indicates the stretched exponential decay Eq. (2) with the $\beta =1/2$. The fundamental (pumping) period $T_f \simeq 40$d and $T_0=1/f_0 \simeq 159$d. The EPO Hamiltonian chaos characteristics are similar to those of the NAO (cf. Figs. 6 and 4). \\

  Another north-south dipole of anomalies is based on the regions 50-70N; 140E-150W and 25-40N; 140E-150W and is known correspondingly as Western Pacific Oscillation (WPO) \cite{bl},\cite{kal},\cite{wpo} (cf. Figures 5 and 7).
  
  In the positive phase of the WPO pressures, heights and temperatures are higher to the south and lower to the north while the negative phase corresponds to an opposite pattern (cf. Fig. 7).

\begin{figure} \vspace{-1cm}\centering
\epsfig{width=.45\textwidth,file=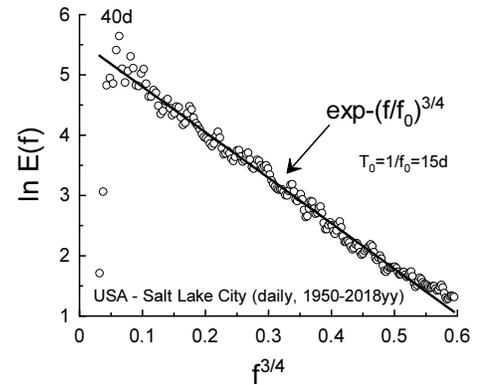} \vspace{-4cm}
\caption{Power spectrum corresponding to the surface air temperature fluctuations for Salt Lake City (the intermountain region of the North America).} 
\end{figure}
\begin{figure} \vspace{-1.2cm}\centering
\epsfig{width=.45\textwidth,file=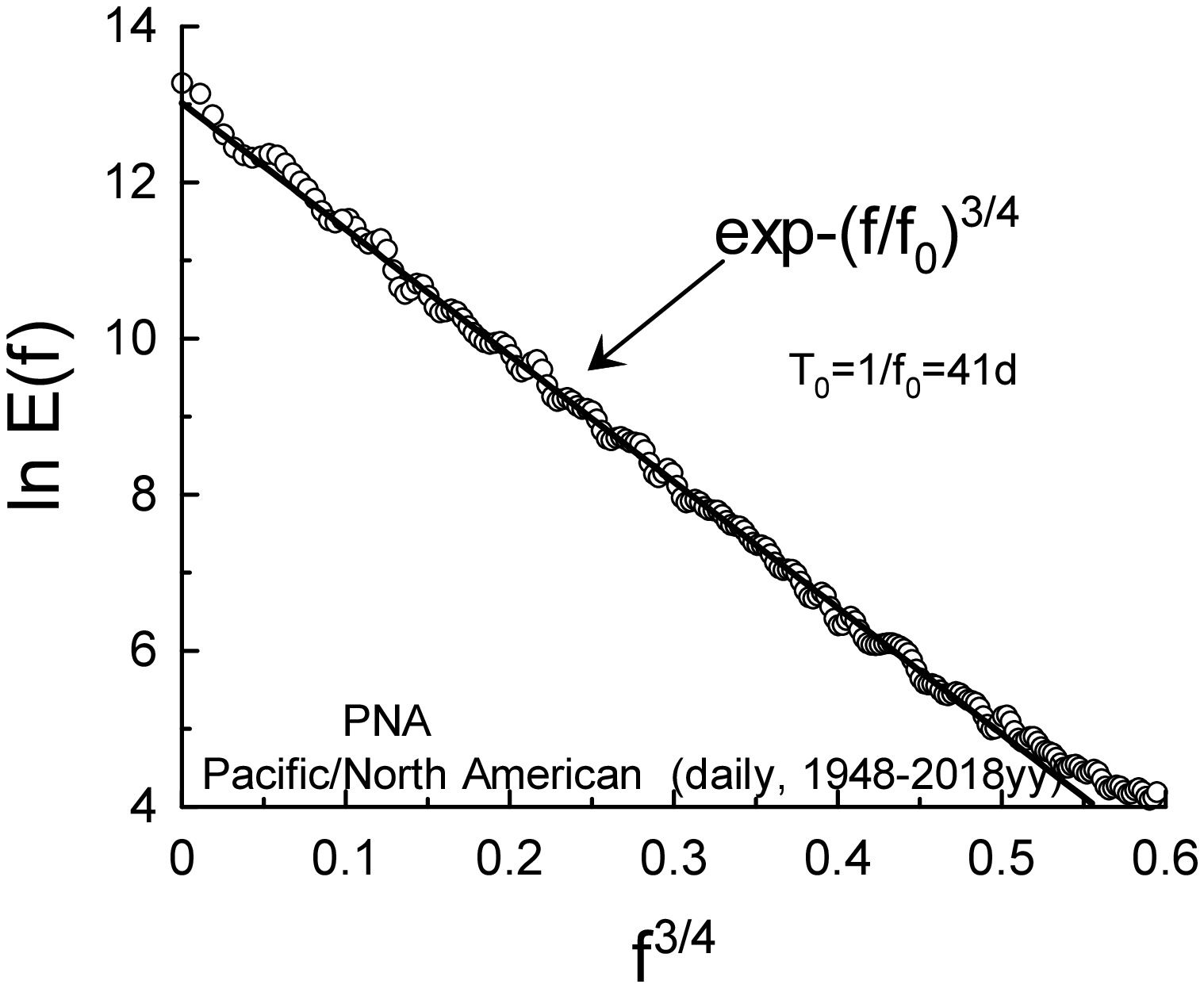} \vspace{-4.3cm}
\caption{Power spectrum for the {\it raw} daily time series corresponding to the PNA index for 1948-2018yy period (the data were taken from the site Ref. \cite{pna}.} 
\end{figure}
\begin{figure} \vspace{-0.5cm}\centering
\epsfig{width=.45\textwidth,file=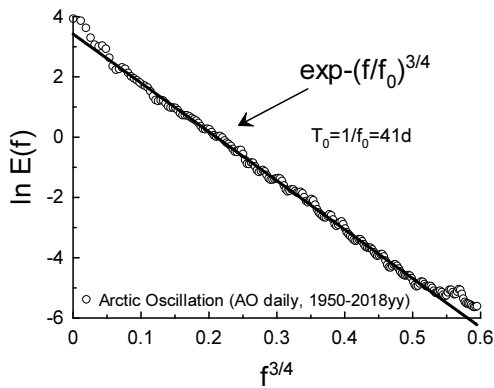} \vspace{-4.1cm}
\caption{As in Fig. 12 but for the AO index for 1950-2018yy period (the data were taken from the site Ref. \cite{ao}).} 
\end{figure}
       
  Figure 8 shows power spectrum computed for the Haar wavelet detrended daily time series of the WPO index for the 1948-2018yy period (the data were taken from the site Ref. \cite{wpo}). The straight line in the figure (the best fit) indicates the stretched exponential decay Eq. (2) with the $\beta =1/2$. The fundamental (pumping) period $T_f \simeq 44$d and $T_0=1/f_0 \simeq 158$d.     \\
   
    It should be noted that the $T_0 \simeq 159$d is practically the same for all the north-south dipole teleconnections computed by the second method, and is the same as for the wavelet detrended global temperature fluctuations (land) \cite{b1}.
 
\section{ The Pacific/North American pattern}

  The Pacific/North American pattern (PNA) is a quadrupole: [(15-25N, 180-140W)-(40-50N, 180-140W)+(45-60N, 125W-105W)-(25-35N, 90W-70W)] (cf. Fig. 9) \cite{pna}. It is also different from the previous cases because it includes land (the intermountain region of North America) as one of its centers. This difference results in a difference in the spectrum as one can see in Fig. 10. Figure 10 shows power spectrum computed for the Haar wavelet detrended daily time series of the PNA index for the 1948-2018yy period (the data were taken from the site Ref. \cite{pna}). The straight line in the figure (the best fit) indicates the stretched exponential decay Eq. (2) with the $\beta =3/4$. The fundamental (pumping) period $T_f \simeq 44$d and $T_0=1/f_0 \simeq 33$d.  \\
  
  Figure 11 shows (for comparison) power spectrum for the wavelet regression detrended daily time series corresponding to the surface air temperature fluctuations for the Salt Lake City (see the Ref. \cite{b1}). This geographical site belongs to the intermountain region of the North America, which is one of the poles of the PNA quadrupole - Fig. 9 (cf. Figs. 11 and 10: $\beta = 3/4$ for the both spectra).
  
\section{PNA, Arctic Oscillation and Greenland blocking phenomenon}
  
   A chaotic spectral affinity of the PNA pattern to the Arctic oscillation is of especial interest. Figure 12 shows the power spectrum of the {\it raw} PNA index (i.e. {\it without} removing the annual and seasonal modes, daily time series were taken from the site Ref. \cite{pna}). Fig. 13 shows 
analogous power spectrum for the raw daily time series corresponding to the Arctic Oscillation index (daily time series were taken from the site Ref. \cite{ao}). One can see a very strong similarity between the spectra shown in the Fig. 12 and Fig. 13 in the appropriately choosen scales (corresponding to the Hamiltonian distributed chaos Eq. (2) with $\beta =3/4$ and $T_0=1/f_0 \simeq 41$d). \\

     Let us recall that the Arctic Oscillation (AO) is a primary annular mode of atmospheric circulation in the Northern Hemisphere  \cite{tw} (see for a review Ref. \cite{w}). This pattern represents a pressure gradient between the polar and subpolar regions and it is characterized by strong circulating winds around the Arctic's perimeter. When the AO index is in its positive phase, these perimeter winds constrain colder air to the polar region. When the index is in its negative phase the winds' confinement is weaker and the colder air masses penetrate deeper into subpolar and the mid-latitudes regions. \\

   The Greenland blocking phenomenon (a quasi-stationary large-scale pattern in the atmospheric pressure field - Fig. 14) is also interesting in this context. Greenland is physiographically a part of the North America and it penetrates deeply into Arctic circle. It is known that the Greenland blocking is important for mid-latitude climate because it diverts the atmospheric polar jet streams southwards or northwards depending of its state (see recent Ref. \cite{han2} and references therein).  \\
   
   The     Greenland Blocking Index (GBI) is defined as a 500mb geopotential height area averaged [60-80N, 280-340E] - Figure 14 (the daily averaged NCEP/NCAR reanalysis was used for the index computations, see for more details Ref. \cite{gbi} and for a review Refs. \cite{han2},\cite{han}).\\
   
   Figure 15 shows power spectrum computed for the raw daily time series of the GBI index for the 1948-2015yy period (the data were taken from the site Ref. \cite{gbi}). The straight line in the figure (the best fit) indicates the stretched exponential decay Eq. (2) with the $\beta =3/4$ and $T_0=1/f_0 \simeq 41$d (cf. Figs. 15, 13 and 12).\\

  Finally, it should be noted that the fundamental period $T_f$ for the north-south teleconnections is between 40 and 44 days (the fundamental periods are indicated by the vertical arrows in the Figs. 4,6,8,10), and the $T_0=1/f_0 \simeq 41$d for the spectra corresponding to the raw time series (Figs. 12,13 and 15). For the Northern Hemisphere extratropics the atmospheric dynamics periods about 40 days are well known observationally (cf. Refs. \cite{b1},\cite{mag}-\cite{cun} and references therein). 
  
\begin{figure} \vspace{-1cm}\hspace*{-0.2cm}\centering
\epsfig{width=.5\textwidth,file=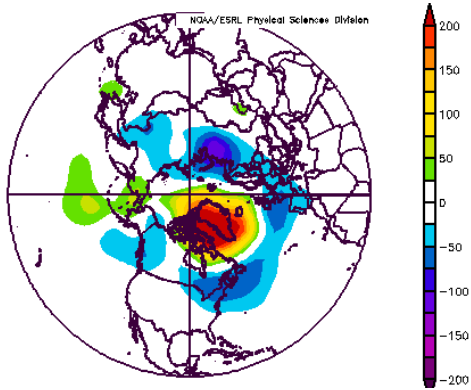} \vspace{-5.5cm}
\caption{Map of the Greenland blocking at 500mb geopotential height \cite{gbi}.} 
\end{figure}
\begin{figure} \vspace{-1cm}\centering
\epsfig{width=.45\textwidth,file=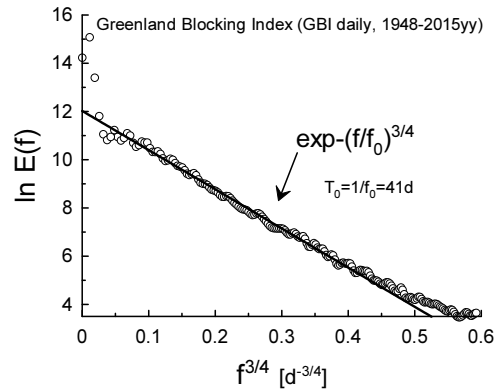} \vspace{-3.9cm}
\caption{Power spectrum for the {\it raw} daily time series corresponding to the GBI index for 1948-2015yy period (the data were taken from the site Ref. \cite{gbi}).} 
\end{figure}

\section{Acknowledgement}

   I acknowledge use of the data provided by the Zenodo database (CERN), the Climate Prediction Center and the Earth system research laboratory (NOAA, USA).

\end{document}